# Ferro-deformation at the nuclear system with protons, Z = 20 and neutrons, N = 40: $^{60}$Ca

Chang-Bum Moon*

*Hoseo University, Chung-Nam 336-795, Korea*

(April 18, 2016)

We present a possibility that the system with Z = 20, N = 40, $^{60}$Ca, has a large deformation, even though it has both proton and neutron magic numbers, symbolizing a spherical nucleus. This large deformation corresponds to the so-called ferro-deformation that occurs at the particular critical points over the nuclear chart. By comparisons with the ferra-deformation at the critical point Z = 40, N = 64 [arXiv:1604.02786], we draw a conclusion that shape phase transitions should occur at Z = 18 or 20 when N = 36 to 38, which leads to a ferro-deformation at the critical points of Z = 18 or 20, N = 40; $^{58}$Ar, $^{60}$Ca. We explain the shape phase transition in terms of isospin dependent spin-orbital interactions between neutrons in the $f_{5/2}$ orbital and protons in the $f_{7/2}$ orbital. We find a universal behavior over the nuclear chart for yielding the ferro-deformation such that; Z = 64, N = 104, Z = 40, N = 64, and Z = 20, N = 40, respectively. This feature is linked to concept of the neutron (ν) − proton (π) interaction in spin-orbital couplings such that; $\nu h_{9/2}$ − $\pi h_{11/2}$, $\nu g_{7/2}$ − $\pi g_{9/2}$, and $\nu f_{5/2}$ − $\pi f_{7/2}$, respectively. It is suggested that triple shape coexistence would be possible in the Z = 20, N = 36, $^{56}$Ca, and the Z = 18, N = 36, $^{54}$Ar, where the ferro-deformation is expected to be built on the second $0^+$ state. The predicted level schemes for $^{52}$Ar, $^{54}$Ar, $^{56}$Ar, $^{56}$Ca, $^{58}$Ca, and $^{60}$Ca are presented.

*cbmoon@hoseo.edu

## 1. Introduction

In the nuclear many-body quantum systems, the so-called magic number is famous. This magic number corresponds to the nucleon number at a large shell closure that exhibits extra stability compared to neighboring nuclei. By introducing a strong spin-orbital coupling term into the single particle potential [1, 2], the magic numbers could be reproduced such as, as shown in Fig. 1(a); 8, 20, 28, 50, 82, and 126. In the practical experimental circumstances, based on the first $2^+$ excited states in Fig. 1(b), we can see such a large energy gap at N = 20 and 28. Dynamical shell structure changes along the lines of both protons (Z) and neutrons (N) = 20, 28, and 40 have been main subjects of the studies in nuclear physics, experimentally and theoretically. Although the number 40 is not generally accepted as a magic number, we include $^{90}$Zr and $^{68}$Ni as a doubly-magic nucleus [3]. It should be noticed that the numbers 8, 20, and 40 are also shell gap numbers built on the harmonic oscillator potential, as shown in Fig. 1(a).

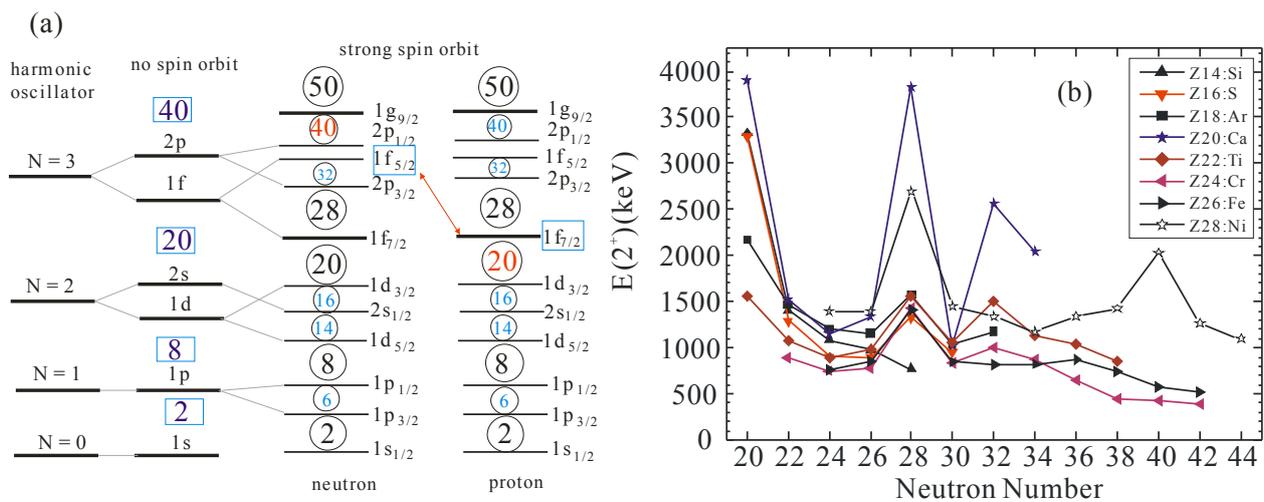

Fig. 1. (a) The single particle energies of a harmonic oscillator potential as a function of the oscillator quantum number N, including the single-particle energies of a Woods-Saxon potential and the level splitting due to the spin-orbital coupling term. The numbers at the energy gaps are the subtotals of the number of nucleons that can occupy each subshell. For discussions, we denote the corresponding spin-orbital doublet, by an arrow, and the critical points of Z = 20 and N = 40 by red color. (b) Systematics for the first $2^+$ excited states in the nuclides over Z = 14 to 28 as a function of neutron numbers. Data are taken from the National Nuclear Data Center [4].

In previous works [5, 6], we demonstrated that a reinforcement driven by proton-neutron interactions in the spin-orbital doublet configuration plays an essential role in giving rise to a huge deformation at particular proton and neutron numbers; Z = 64, N = 104 [5] and Z = 40, N = 64 [6]. We defined this reinforced deformation as the ferro-deformation.

In this work, by comparisons with the systematics of E($2^+$) and R = E($4^+$)/E($2^+$) values in other regions [5, 6], we offer the shape phase transition, shape coexistence, and ferro-deformation for the region, $8 \leq Z \leq 28$ and $20 \leq N \leq 44$. We suggest that the nuclear system with Z = 20, one of the magic numbers in the nuclear many-body systems, has ferro-deformation at N = 40; $^{60}$Ca. This ferro-deformation is explained as being attributed to the isospin dependent spin-orbital interactions between neutrons in the $f_{5/2}$ orbital and protons in the $f_{7/2}$ orbital. For a visual representation of the shape phase transition and ferro-deformation in the nuclei Ar and Ca, with N = 34, 36, 38, and 40, the proposed level schemes will be given.

## 2. A ferro-deformation near Z = 20, N = 40.

Figure 2 illustrates the systematics of the first $2^+$ energy and the excitation energy ratios of the first $2^+$ and $4^+$ states, R = $E(4^+)/E(2^+)$. This R value, as a deformation parameter, provides us with nuclear shape information such as; a spherical shape at R < 2, a vibrator based on the weak spherical at R ~ 2, and a deformed axial rotator at R = 3.3 [7].

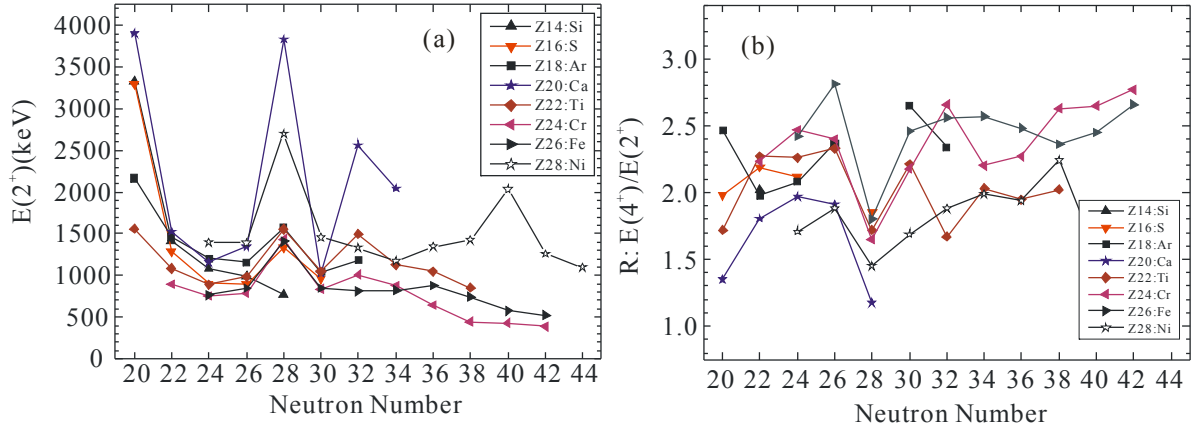

Fig. 2. Systematic plots of (a) $E(2^+)$ and (b) R, $E(4^+)/E(2^+)$, values as a function of neutron numbers for the nuclei within the Z = 14 and 28 space. Data are taken from the National Nuclear Data Center [4].

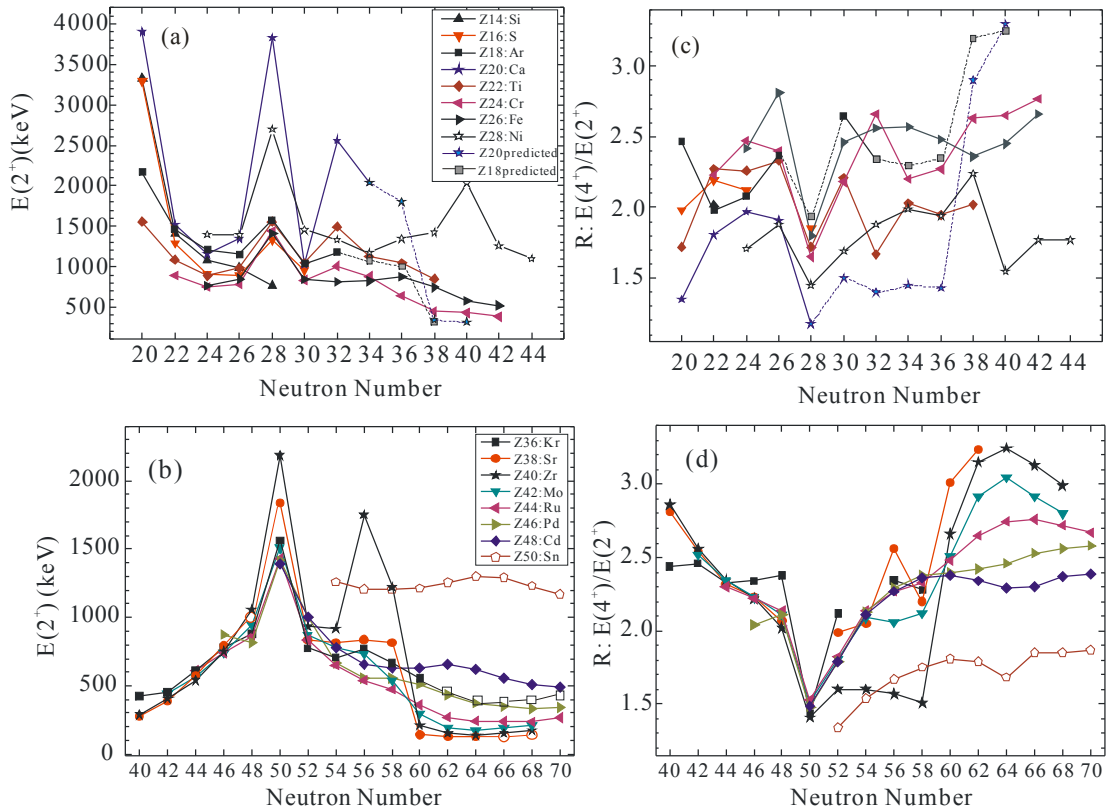

Fig. 3. Systematic plots for $E(2^+)$ and R, $E(4^+)/E(2^+)$, values as a function of neutron numbers for the nuclei; (a) and (c), Z = 14 to 28, and (b) and (d), the Z = 36 to 50 space [6]. The dotted lines in (a) and (c) are the predicted values as obtained by systematic arguments. See the text for more details and Fig. 4.

In the region within both Z = 14 and 28 and N = 20 and 44, as far as the current nuclear data at NNDC are involved,

we find that there is no sign revealing a large deformation, indicating R > 3, along with a sudden shape phase transition. However, by a comparison of the systematics with those observed in other regions, for instance, the Z = 28 to 40 space, we are able to get some insights into signaling a shape phase transition indicating a spherical shape to a deformed shape. Figure 3 shows the systematics of the E(2⁺) and the R values in the two regimes; one is in the vicinity of Z = 20, N = 40 and the other of Z = 40, N = 64. From the comparisons of both parameters, we can deduce a critical point that would take place a shape phase transition. Firstly, we notice that a phase transition occurs at Z = 36 to 38 as shown in Fig. 3(b) and (d). Secondly, by utilizing concepts of the pseudo-shell built on the combined subshells, as in the previous works [5, 6], we expect a critical point at which the ferro-deformation occurs.

Table 1. Summary of pseudo-shell configurations between double-shell closures of $8 \leq Z \leq 28$ and $28 \leq N \leq 50$: pseudo-shells, involving subshells, total spins, half-filled nucleon numbers, and distinctive deformation configurations.

| pseudo-shells | combined subshells | pseudo-shell total spins: nucleon capacitance | half(full)-fill summed nucleon number | representative deformation configurations: (Z, N) |
|---|---|---|---|---|
| $J_{ds}(7/2)$ | $1d_{5/2}2s_{1/2}$ | 7/2: 8 | 12 (16) | (12, 20) |
| $J_{dsd}(11/2)$ | $1d_{5/2}2s_{1/2}1d_{3/2}$ | 11/2: 12 | 14 (20) | (14, 20) |
| $J_f(7/2)$ | $1f_{7/2}$ | 7/2: 8 | 24 (28) | (24, 32) |
| $J_{dsf}(19/2)$ | $1d_{5/2}2s_{1/2}1d_{3/2}1f_{7/2}$ | 19/2: 20 | 18 (28) | (18, 38) |
| $J_{pf}(9/2)$ | $2p_{3/2}1f_{5/2}$ | 9/2: 10 | 32 or 34 (38) | (20, 32) |
| $J_{pf}(11/2)$ | $2p_{3/2}1f_{5/2}2p_{1/2}$ | 11/2: 12 | 34 (40) | (20, 34) |
| $J_{pfg}(21/2)$ | $2p_{3/2}1f_{5/2}2p_{1/2}1g_{9/2}$ | 21/2: 22 | 38 or 40 (50) | (18, 38), (20, 40) |

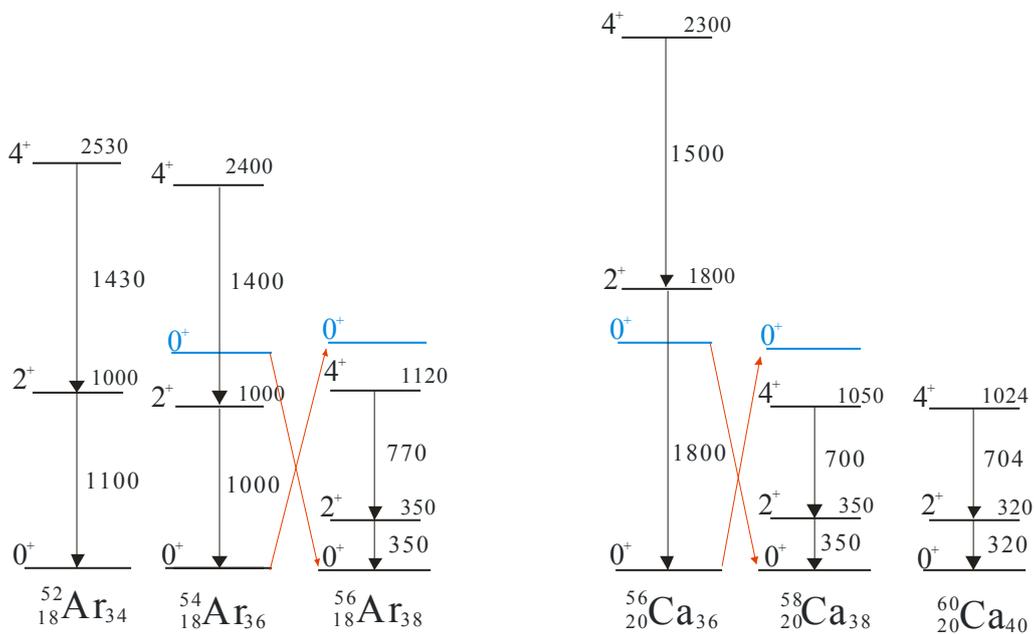

Fig. 4. The proposed level schemes for (a) Z = 18 and (b) Z = 20 with N = 34, 36, 38, and 40. It is noted that an assumed second 0⁺ state is added for indicating possible shape coexistence. We suppose that the uncertainty for the 2⁺ transition energy to the ground state would be about 20 % in each nucleus.

In Table 1, we summarize the pseudo-shell configurations along with the Z, N combinations that yield some distinctive shell structures in the region; $8 \leq Z \leq 28$ and $28 \leq N \leq 50$. For protons on one side, the pseudo-shell, $J_{dsf}(19/2)$ has the capacitance of nucleons 20 while in the neutron region, the $J_{pfg}(21/2)$ has the 22 capacitance. We know that a reinforced deformation takes place when both proton and neutron pseudo-shells are half-filled, which corresponds to Z = 18 and N = 38. As with the previous studies for the ferra-deformation at the critical point Z = 40, N = 64 [6], we expect that a shape phase transition should occur at Z = 18 or 20 when N = 36 to 38. In turn, a ferro-deformation is formed at the critical points of Z = 18 or 20, N = 40. Following this scenario and the systematics based on the E(2$^+$) and R parameters in the region of Z = 40, N = 64, we predict the E(2$^+$) and R values for the Z = 18 and 20 with N = 34, 36, 38, and 40. Our results are shown in Fig. 3(a) and (c), in which the predicted values are plotted as the points with dotted lines. In addition, we present the level schemes based on the predictive values in Fig. 4, as showing a ferro-deformation structure in these nuclei.

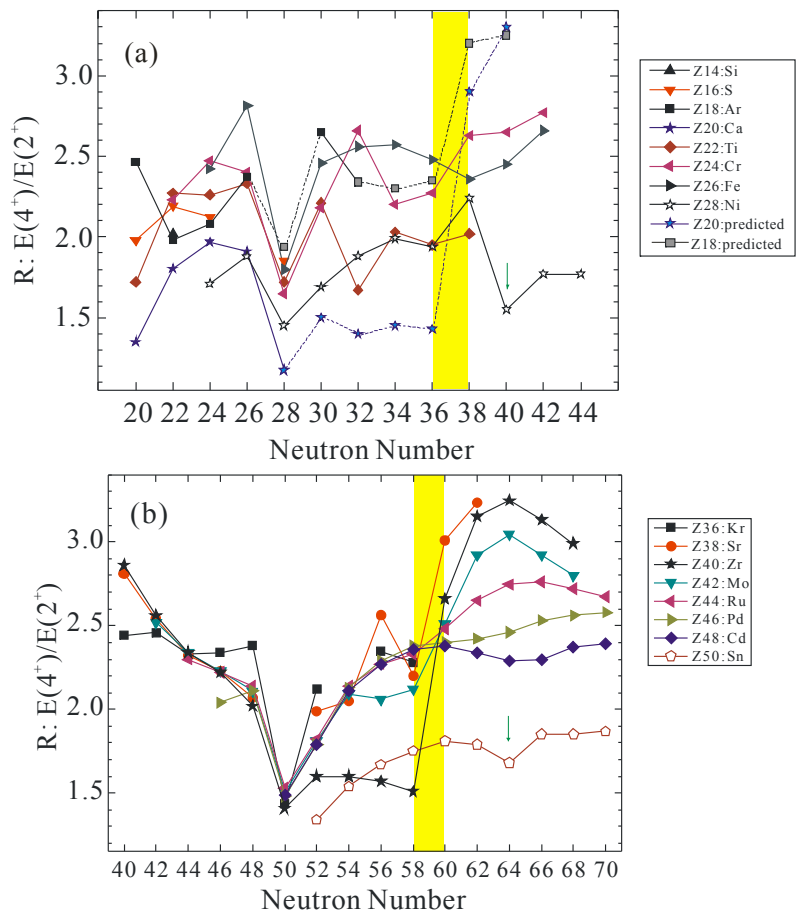

Fig. 5. Plots of the systematics for the R (= E(4$^+$)/E(2$^+$) values as a function of neutron numbers in isotopes between (a) Z = 14 and 28, and (b) Z = 36 and 50 [6]. The shaded regions include the critical points that shape phase transitions take place. The points connected with dotted lines correspond to the expected values by the present work. Notice the point at N = 40 and 64, respectively denoted by an arrow, which indicates a semi-double shell closure.

## 3. The twin magic numbers: Z = 20 and 40.

In order for our prediction of the ferro-deformation at Z = 20 to be feasible, a comparison with those at Z = 40 has to be made. We plot such comparisons of the R values in the two regimes; one is around Z = 20, N = 40 and the other around Z = 40, N = 64, in Fig. 5 for the isotopes as a function of neutrons, and in Fig. 6 for the isotones as a function of protons.

Now we find the very similarity between two regimes. The distinctive features are as follows; first, the feature at Z = 20, N = 40 bears resemblance to that at Z = 40, N = 64. Here we would like to emphasize that the nucleon numbers 64, 40, and 20 contribute to producing a semi-double shell closure. It is also noticed that the numbers 20 and 40 are the shell gap numbers built on the harmonic oscillator. Furthermore, Z = 64 gives rise to the ferro-deformation with N = 104 [5]. Second, we find a universal behavior over the nuclear chart for yielding a ferro-deformation such that; Z = 64, N = 104 [5], Z = 40, N = 64 [6], and Z = 20, Z = 40 in the present work. This feature is closely related to the proton-neutron interactions in spin-orbital doublet such that; $\nu h_{9/2} - \pi h_{11/2}$, $\nu g_{7/2} - \pi g_{9/2}$, and $\nu f_{5/2} - \pi f_{7/2}$, respectively.

We find, furthermore in view of the pseudo-shell configurations, a similar pattern of shape phase transitions between the two proton regimes; $J_{dsf}(19/2)$ and $J_{pfg}(21/2)$ such that; the N =36 and 38 correspond to the N = 58 and 60, and the Z = 18 to 20 correspond to the Z = 38 to 40. From Fig. 6, we also readily find such a similar correlation between Z = 18, 20 and Z = 38, 40. We emphasize that; for Z = 20, the points at N = 20 and 28 characterize a spherical shape indicating the double-magic shell gap while the point at N = 40 has a huge deformed shape, turning out *absence* of such a magic number. For the Z = 40 system, a similar feature is seen at N = 50 and 64.

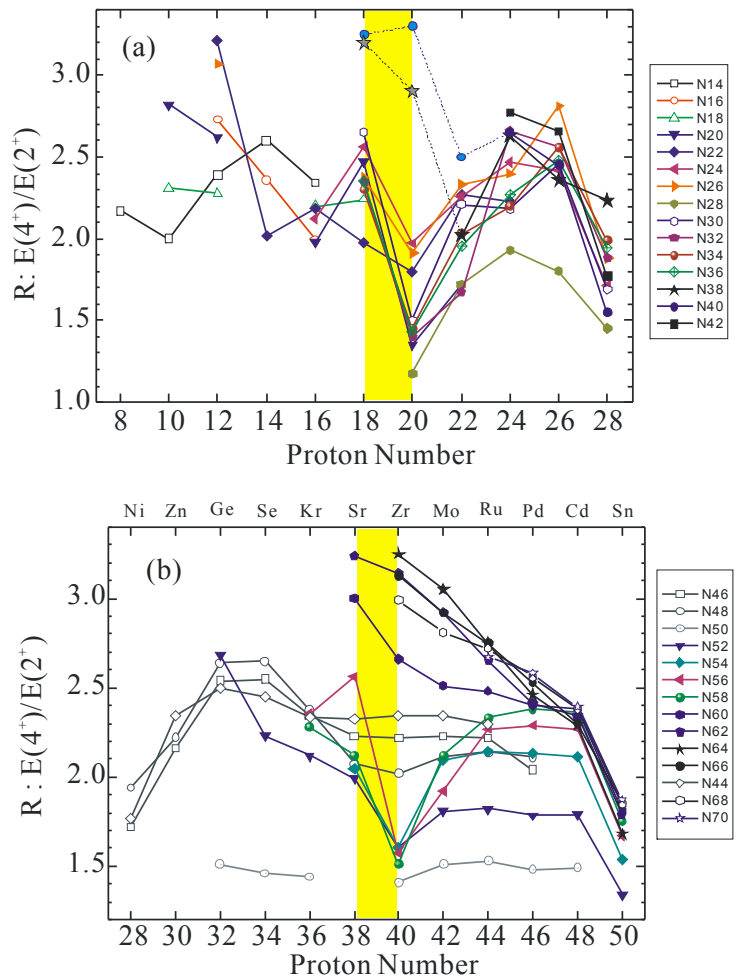

Fig. 6. Systematic plots for R (= E(4$^+$)/E(2$^+$)) values as a function of proton numbers in isotones between (a) N = 14 and 42, and (b) N = 46 and 70 [6]. The shaded regions indicate the critical points occurring shape phase transitions. The points with dotted lines indicate the values predicted in the present work.

If we assumed a path following the cases of the Z = 40 [6] and 64 [5], the points of Z = 20 at N = 30, 32, 34, and 36 generate a similar feature with those of Z = 40 at N = 52, 54, 56, and 58. More surprisingly, such assumed points overlap appreciably with those of Z = 28 at N = 30, 32, 34, and 36, as indicating a typical vibrational structure. According to the above systematic features, we conclude that there is a possibility of triple shape coexistence in $^{56}$Ca, including a ferro-deformation. In a sense, ***the proton numbers 20 and 40 look like a twin, having the very similar structural character.***

## 4. Conclusions

We provided an evidence that the system with Z = 20, N = 40, $^{60}$Ca, has a large deformation, even though it has both proton and neutron magic numbers. This suggestion has been made by a careful comparison with the systematic features, based on the first $2^+$ level energies, E($2^+$), and the ratios of the first $4^+$ level and $2^+$ level energies, R = E($4^+$)/E($2^+$), between two proton regimes; one is $8 \leq Z \leq 28$ and the other is $28 \leq Z \leq 50$. Shape phase transitions including a reinforced deformation were explained in terms of the *pseudo-shell* configurations and the isospin dependent spin-orbital interactions. An important consequence is the distinction between Z = 18, 20, N = 38, 40 and Z = 38, 40, N = 58, 60 for shape phase transitions. The reinforced deformation takes place when Z = 18 or 20 correlates with N = 38 and reaches R = 3.3, over Z = 20, N = 40, yielding the so-called ***ferro-deformation*** [5, 6]. It is also expected that the ferro-deformation would appear as a rotational band built on the second $0^+$ state in $^{56}$Ca and $^{54}$Ar, respectively. Our results provide an outstanding example of shape phase transitions over the nuclear mass chart. We hope that searching for the second $0^+$ and $2^+$ states in $^{56}$Ca and $^{54}$Ar would be made at a current rare isotopes beam facility such as; the Facility for Rare Isotope Beams (FRIB) at MSU in the U.S.A, the Radioactive Isotope Beams Factory (RIBF) at RIKEN in Japan, the Fragment Separator (FRS) at GSI in Germany, and HIE-ISOLDE at CERN in Europe.

**Note added 1**: The present results have been made with phenomenological arguments based on the systematics of experimental observables given by the national nuclear data center, NNDC. Accordingly, many references, possibly related to the present work in the literature, could not be quoted because of the referred data from NNDC.

**Note added 2**: Any abbreviation is avoided since it makes the readers confusing like a jargon.

## References

[1] M. G. Mayer, Phys. Rev. **75**, 1969 (1949).

[2] O. Haxel, J. H. D. Jensen, and H. E. Suess, Phys. Rev. **75**, 1766 (1949).

[3] Chang-Bum Moon, AIP Adv. **4**, 041001 (2014).

[4] National Nuclear Data Center, Brookhaven National Laboratory, http:// www.nndc.bnl.gov/ (March 2016).

[5] Chang-Bum Moon, **arXiv**:1604.01017 (2016).

[6] Chang-Bum Moon, **arXiv**:1604.02786 (2016).

[7] K. S. Krane, Introductory Nuclear Physics (John Wiely & Sons, New York, 1988).